\documentclass[aps,prd,amsfonts,preprint,eqsecnum]{revtex4}

\usepackage{graphics,epsfig}

\begin{document}



\preprint{JLAB-THY-03-02; WM-03-101}

\title{ Angular Conditions, Relations between Breit and Light-Front 
Frames, and Subleading Power Corrections\\ 
}

\author{ Carl E. Carlson$^{ab}$}
\email[]{carlson@physics.wm.edu}

\author{Chueng-Ryong Ji$^{c}$}
\email[]{crji@unity.ncsu.edu}

\affiliation{
$^{a}$Thomas Jefferson National Accelerator Facility,
12000 Jefferson Avenue, Newport News, VA 23606\\
$^{b}$Nuclear and Particle Theory Group, Physics Department,
College of William and Mary,
Williamsburg, Virginia 23187-8795\\
$^{c}$Department of Physics,
North Carolina State University,
Raleigh, North Carolina 27695-8202}

\date{printed \today}



\begin{abstract}
We analyze the current matrix elements in the general collinear (Breit)
frames and find the relation between the ordinary (or canonical)  
helicity amplitudes and the light-front helicity amplitudes. 
Using the conservation of angular momentum, we derive a general angular 
condition which should be satisfied by the light-front helicity amplitudes
for any spin system. In addition, we obtain the light-front parity 
and time-reversal relations for the light-front helicity amplitudes.
Applying these relations to the spin-1 form factor analysis, we note 
that the general angular condition relating the five helicity 
amplitudes is reduced to the usual angular condition relating the four 
helicity amplitudes due to the light-front time-reversal condition. 
We make some comments on the consequences of the angular condition for the analysis of the high-$Q^2$ deuteron electromagnetic form factors, and we  further apply the general angular condition to the electromagnetic 
transition between spin-1/2 and spin-3/2 systems and find a relation 
useful for the analysis of the N-$\Delta$ transition form factors.
We also discuss the scaling law and the subleading power corrections in 
the Breit and light-front frames. 

\end{abstract}

\maketitle

\newpage



\section{Introduction}


A relativistic treatment is one of the essential ingredients that
should be incorporated in describing hadronic systems. The hadrons have 
an intrinsically relativistic nature since the quantum chromodynamics
(QCD) governing the quarks and gluons inside the hadrons has 
{\it a priori} a
strong interaction coupling and the characteristic momenta of quarks
and gluons are of the same order, or even very much larger, than the
masses of the particles involved. It has also been realized that a
parametrization of nuclear reactions in terms of non-relativistic wave
functions must fail. In principle, a manifestly covariant framework
such as the Bethe-Salpeter approach and its covariant equivalents can
be taken for the description of hadrons. However, in practice, such
tools are intractable because of the relative time dependence and the
difficulty of systematically including higher order kernels. A
different and more intuitive framework is the relativistic Hamiltonian
approach. With the recent advances in the Hamiltonian renormalization
program, a promising technique to impose the relativistic treatment of
hadrons appears to be
light-front dynamics (LFD), in which a
Fock-space expansion of bound states is made at equal light-front time
$\tau = t+z/c$. The reasons that make LFD so attractive to solve
bound-state problems in field theory make it also useful for a
relativistic description of nuclear systems.

Light-front quantization \cite{Dirac,Steinhardt} has already been
applied successfully in the context of current algebra \cite{lcca} and
the parton model \cite{lcparton} in the past. For the analysis of
exclusive processes involving hadrons, the framework of light-front
(LF) quantization~\cite{BPP} is also one of the most popular
formulations. In particular, the 
light-front or Drell-Yan-West ($q^+=q^0+q^3=0$) frame
has been extensively used in the calculation of various electroweak
form factors and decay processes~\cite{Ja2,CJ1,Kaon}. In this
frame~\cite{DYW}, one can derive a first-principle formulation for the
exclusive amplitudes by choosing judiciously the component of the
light-front current. As an example, only the parton-number-conserving
(valence) Fock state contribution is needed in $q^+=0$ frame when a
``good" component of the current, $J^+$ or ${\bf J}_{\perp}=(J_x,J_y)$,
is used for the spacelike electromagnetic form factor calculation of
pseudoscalar mesons. One doesn't need to suffer from complicated vacuum
fluctuations in the equal-$\tau = t+z/c$ formulation due to the
rational dispersion relation. The zero-mode contribution may also be
avoided in Drell-Yan-West (DYW) frame by using the plus component of
current~\cite{Ji1}. The perturbative QCD (PQCD) factorization theorem
for the exclusive amplitudes at asymptotically large momentum transfer
can also be proved in 
LFD formulated in the DYW frame.

However, caution is needed in applying the established Drell-Yan-West
formalism to other frames because the current components do mix under
the transformation of the reference-frame~\cite{chad}. Especially, for
the spin systems, the light-front helicity states are in general
different from the ordinary (or canonical) helicity states which may be
more appropriate degrees of freedom to discuss the angular momentum
conservation. As the spin of the system becomes larger, the number of
current matrix elements gets larger than the number of physical form
factors and the conditions that the current matrix elements must
satisfy are essential to test the underlying theoretical model for the
hadrons. Thus, it is crucial to find the relations between the ordinary
helicity amplitudes and the light-front helicity amplitudes in the
frame that they are computed.

In this work, we use the general collinear frames which cover both
Breit and target-rest frames to find the relations between the ordinary
helicity amplitudes and the light-front helicity amplitudes. Using the
conservation of angular momentum, we derive a general angular condition
which can be applied for any spin system. The relations among the
light-front helicity amplitudes are further constrained by the
light-front parity and time-reversal consideration. For example, the
spin-1 form factor analysis requires in general nine light-front
helicity amplitudes although there are only three physical form
factors. Thus, there must be six conditions for the helicity
amplitudes. Using the light-front parity relation, one can reduce the
number of helicity amplitudes down to five. The general angular
condition 
gives one relation among the five light-front helicity 
amplitudes, leaving four of them independent. 
One more relation comes by applying the light-front time-reversal
relation, also having the effect that the general anglar conditon can be reduced to the usual angular condition relating only four helicity ampituds.  Consequently, only three helicity amplitudes are independent
each other, as it should be because there are only three physical form
factors in spin-1 systems. We also 
apply the general angular condition to the electromagnetic transition between spin-1/2 and spin-3/2 systems and 
find the relation among the helicity amplitudes
that can be used in the analysis of the N-$\Delta$ transition. In
particular, the angular condition provides a strong constraint to the
N-$\Delta$ transition indicating that the suppression of the helicity
flip amplitude with respect to the helicity non-flip amplitude for the
momentum transfer $Q$ in PQCD is in an order of $m/Q$ or $M/Q$ rather
than $\Lambda_{QCD}/Q$, where both nucleon mass $m$ and delta mass $M$
are much larger than the QCD scale $\Lambda_{QCD}$. Thus, one may
expect that the applicability of leading PQCD could be postponed to a
larger $Q^2$ region than one may naively anticipate from 
leading PQCD. The same consideration can apply for the deuteron form
factor analysis from the spin-1 angular condition. This work presents
further discussions on the scaling law and the subleading power
corrections in the Breit and light-front frames.

The paper is organized as follows. In the next section (Section II), we
present the derivation of transformation laws between the ordinary
helicity amplitudes and the light-front helicity amplitudes and obtain
a general angular condition on the current matrix elements using the
rotational covariance of the current operator. Since we start from the
definition of states in a general collinear frame, our derivation may
be more physically transparent than any other formal derivation. 
In Section III, we present the light-front discrete symmetries and derive the parity and time-reversal rules for the helicity amplitudes. 
In Section IV, we discuss the consequences from these
findings of the general angular condition and the light-front discrete
symmetry relations. The reduction of number of independent helicity 
amplitudes is shown for a few example spin systems. The current matrix 
elements of the spin-1 system and the spin-1/2 to
spin-3/2 transition are shown as explicit examples. 
The subleading power corrections are obtained from
the general angular condition and the scaling laws are derived for the
ordinary Breit frame helicity amplitudes and the light-front helicity
amplitudes. Summary and conclusion follow in Section V.


\section{Frame Relations and General Angular Condition} \label{relatons}


Our subject is relations among matrix elements or helicity amplitudes 
for the process $\gamma^*(q) + h(p) \rightarrow h'(p')$, where 
$\gamma^*$ is an off-shell photon of momentum $q$, and $h$ and $h'$ are 
hadrons with momenta $p$ and $p'$, respectively.  
(Results will be easily extendable for other incoming vector bosons.)  

Calculations may be done in the light-front frame, which is 
characterized by having $q^+ \equiv q^0 + q^3 = 0$, and may be done in 
the Breit frame, which is  characterized by having the  photon and 
hadron 3-momenta along a single line.  Each frame has its
advantages.  
In the light front frame with $q^+ = 0$, and for matrix elements of the 
current component $J^+$, the photon only couples to forward moving 
constituents (quarks) of the hadrons and never produces a 
quark-antiquark pair.  Thus one only needs wave functions for
hadrons turning into constituents going forward in (light-front) time, 
and can develop a simple parton picture of the interaction.  On the 
other hand, the Breit frame, being a collinear frame, makes it easy to 
add up the helicities of the incoming and outgoing particles and to 
count the number of independent non-zero amplitudes.

By transforming efficiently back and forth one can realize the 
advantages of both frames.  Hence our first goal in this section will be 
to find the relations between the light-front and Breit frame helicity 
amplitudes, and then to use those relations to derive in a transparent 
way the general relation among the light-front amplitudes that is 
usually referred to as the ``angular condition.''


\subsection{Relations among helicity amplitudes}


Connecting light-front and Breit helicity amplitudes is facilitated by 
finding frames that are both simultaneously realized.
One excellent and easy 
example is the particular light-front frame where the target is at 
rest.  This is also a Breit frame, since with the target 3-momentum 
zero, the remaining momenta must lie along the same line.  We are 
perhaps extending the idea of a Breit frame, but are doing so in a way 
that leaves invariant the Breit frame helicity amplitude.  That is, one 
normally thinks of a Breit frame as one where the incoming and outgoing 
hadron have oppositely directed momenta,  along the same line.  
Sometimes one specifies that the line is the $z$-axis.  However, since 
helicities are unaffected by rotations~\cite{jacobwick}, one can choose any 
line at all.  Further, helicities are unaffected by collinear 
boosts~\cite{jacobwick} that don't change the particle's momentum 
direction.   One can also boost along the direction of motion until one 
of the hadrons is at rest, provided one defines the positive helicity 
direction for the particle at rest to be parallel to the momentum the 
particle would have in conventional Breit frame.  With this natural 
helicity direction choice, the Breit frame  helicity amplitude in a 
target rest frame is (if we use 
relativistic normalization conventions, as we shall always do) precisely 
the same as  the Breit frame amplitude in a conventional Breit frame 
with the same helicity labels.

Thus, the light-front frame with the target at rest is both a Breit 
frame and a light-front frame.  There is a continuum of such frames.  
Another useful example is a Breit frame with the incoming and outgoing 
hadrons moving in the negative and positive $x$-directions, adjusted to 
have equal incoming and outgoing energies.  In this case, $q^0$ and 
$q^3$ are individually zero, so that $q^+=0$ and we have also a 
light-front frame.

Even in a frame that is simultaneously light-front and Breit, the 
connection between the two types of helicity amplitudes can be a bit 
involved.  This is because the definitions of the light-front and 
ordinary helicity states are not the same and the general conversion 
between them for a moving state involves a rotation by an 
angle that is not trivial to determine.  

Our plan will be to use the rest of this subsection to define our 
notation, state the main result for the light-front to Breit and 
vice-versa helicity amplitude conversion formulas, and show how one 
obtains the general angular condition from this result. 
Then in the next subsection, we will give the details of the derivation.

For light-front amplitudes one uses light-front helicity states, which 
for a momentum $p$ are defined by taking a state at rest with the spin 
projection along the $z$-direction equal to the desired helicity, then 
boosting in the $z$-direction to get the desired $p^+$, and then doing a 
light-front transverse boost to get the 
desired transverse momentum $p_\perp$.  We call this state 
\begin{equation}
	| p,\lambda \rangle_L  \ ,
	\end{equation}

\noindent and it is defined by formula in the next subsection.	The spin 
of the particle, $j$, is understood but not usually written, $\lambda$ 
is the light-front helicity of the particle, and the normalization is 
\begin{eqnarray}
	_L\langle p_2,\lambda_2 | p_1, \lambda_1 \rangle_L = 
		(2\pi)^3  2p_1^+  
		\delta(p_1^+ -p_2^+) \delta^2( p_{2\perp}-p_{1\perp}) 
		\delta_{\lambda_1 \lambda_2}     \ .     
	\end{eqnarray}

\noindent The light-front helicity amplitude $G_L$ is a matrix element 
of the electromagnetic current $J^\nu$ given by 
\begin{equation}
	G^\nu_{L\lambda'\lambda} = \ 
		_L\langle p',\lambda' | J^\nu | p, \lambda \rangle_L  \ .
	\end{equation}

In the Breit frame, we use ordinary helicity states, which are defined 
by starting with a state at rest having a spin projection along the 
$z$-direction equal to the desired helicity, then boosting in the 
$z$-direction to get the desired $|\vec p|$, and then rotating to get 
the momentum and spin projection in the desired direction.  (We shall 
generally keep our momenta in the $x$-$z$ plane, so we do not need to 
worry about the distinction between, for example, the 
Jacob-Wick~\cite{jacobwick} helicity states and the somewhat later Wick 
states~\cite{wick62}.)  The state will be denoted, 
\begin{equation}
	| p,\mu \rangle_B  \ ,
	\end{equation}
	
\noindent where $\mu$ is the helicity or spin projection in the 
direction of motion, and the subscript ``$B$'' reminds us which frame we 
use these states in.  Except for momenta directly along the $z$-axis, 
the light-front helicity and regular helicity states are not the 
same, but if the 4-momenta are the same they can be related by a  
rotation.  The Breit frame helicity amplitude $G_B$ is, 
\begin{equation}
	G^\nu_{B\mu'\mu} = \ 
		_B\langle p',\mu' | J^\nu | p, \mu \rangle_B  \ ,
	\end{equation}

\noindent  where $J^\nu$ is the same electromagnetic current.

A main result is the relation between $G_L$ and $G_B$, which is
\begin{equation}                  \label{main}
	G^\nu_{B\mu'\mu} =  d^{j'}_{\lambda'\mu'}(-\theta') \ 
		G^\nu_{L\lambda'\lambda} \ d^j_{\mu\lambda}(\theta)  \ .
	\end{equation}
	
\noindent A sum on repeated helicity indices is implied. The 
$d$-functions are the usual representations of rotations about the 
$y$-axis for particles whose spins are given by the superscript.  The 
angles are given by 
\begin{equation}            \label{thetahalf}
	\tan{\theta\over 2} = {Q_+ Q_- - Q^2 - M^2 + m^2\over 2mQ }
	\end{equation}
	
\noindent and
\begin{equation}             \label{thetaprime}
	\tan{\theta'\over 2} =  - {Q_+ Q_- - Q^2 + M^2 - m^2\over 2MQ } \ ,
	\end{equation}
	
\noindent where $m$ is the mass of the incoming hadron, $M$ is the mass  
of the outgoing hadron, $Q = \sqrt{Q^2}$, 
\begin{equation}
	Q^2 = -q^2 = - (p-p')^2 \ , 
	\end{equation}

\noindent and
\begin{equation}
	Q_\pm = \left( Q^2 + (M \pm m)^2 \right)^{1/2}  \ ,
	\end{equation}

\noindent and we assume that $q^2$ is spacelike (negative, in our 
metric).  For the elastic case, $M=m$, the angles $\theta$ and 
$-\theta'$ are the same.  (It may seem peculiar to have a minus sign 
inserted twice, as it appears
in Eqs.~(\ref{main}) and~(\ref{thetaprime}), 
but it will seem more sensible when one sees how the 
angles arise, in the next subsection.)

In the Breit frame, since it is collinear, the sum of spin projections 
along the direction of motion must be conserved, so that if 
$\lambda_\gamma$ is the helicity of the photon, 
\begin{equation}
	\lambda_\gamma = \mu + \mu'  \ .
	\end{equation}
	
\noindent Even if the photon is off-shell, it cannot have more than one 
unit of helicity in magnitude.  Hence there is a constraint on the Breit 
frame helicity amplitude, 
\begin{equation}
	G^\nu_{B\mu'\mu} = 0 \quad {\rm if} \quad |\mu'+\mu| \ge 2  \ .
	\end{equation}

\noindent This induces a constraint on the light front amplitudes, and 
this constraint is what is called the angular condition, given generally 
by~\cite{capstick95} 
\begin{equation}
	d^{j\prime}_{\lambda'\mu'}(-\theta') \ 
		G^\nu_{L\lambda'\lambda} \ d^j_{\mu\lambda}(\theta) = 0
		\quad {\rm if} \quad |\mu' + \mu| \ge 2  \ ,
	\end{equation}
	
\noindent and most often applied when the Lorentz index $\nu$ is $+$.  
One sees that the result follows from angular momentum conservation and 
the limited helicity of the photon.  We will check in 
section~\ref{consequences} that upon expressing the  $d$-functions in 
terms of $Q^2$ and mass, one obtains the known angular condition for 
electron-deuteron elastic scattering, and  will we also obtain in terms 
of masses and $Q^2$ the angular condition for the N-$\Delta(1232)$ 
electromagnetic transition.


\subsection{Deriving the light-front to Breit relation}


We gave two examples of frames that were simultaneously Breit and 
light-front frames.  It turns out that half the work we need to do is 
very easy in one of these frames, and that visualizing one ensuing 
equality is quite easy in the other.

Clearly, one can write
\begin{equation}
	G^\nu_{B\mu'\mu} =  
		\ _B\langle p',\mu'| p',\lambda' \rangle_L \ 
		G^\nu_{L\lambda'\lambda} \ \,
		_L\langle p,\lambda| p,\mu \rangle_B  \ ,
	\end{equation}

\noindent so that the problem reduces to finding the overlaps of the 
light-front helicity and ordinary helicity amplitudes.  

We shall start using the light-front frame with the target at rest.  For 
the initial state, being at rest, the light-front helicity state is 
identical to the state with spin quantized in the positive $z$-direction,
\begin{equation}
	| p,\lambda \rangle_L = | rest,\lambda \rangle_z  \ . 
	\end{equation}


\begin{figure}

\epsfig{figure=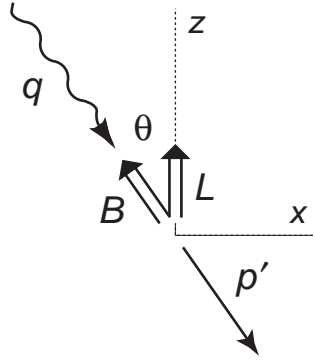}

\caption{Photon with $q^+ = 0$ absorbed on particle at rest.  Two 
choices for the target spin axis are indicated.  In the Breit frame 
($B$), the helicity state positive direction is opposite to the 
direction of the entering photon.  The light-front state 
($L$), in this case, is identical to the rest state quantized along the 
positive $z$-axis.  The angle $\theta$ between the photon 3-momentum 
direction and the (negative) $z$-axis is also the angle between the two 
choices of spin quantization axis.}

\label{one}

\end{figure}


\noindent  The helicity state, however, should be quantized along a 
direction antiparallel to the momentum of the entering photon; see 
Fig.~\ref{one}.  The photon four-momentum is 
\begin{equation}
	q = (q^+, q^-, q_\perp) = (0, {Q^2 + M^2 - m^2 \over m}, Q ) \ ,
	\end{equation}
	
\noindent and it makes an angle $\theta$ with the $z$-axis, where
\begin{equation}
	\tan \theta = { 2 m Q \over Q^2 + M^2 - m^2 } \ ,
	\end{equation}
	
\noindent equivalent to the half-angle version given earlier, 
Eq.~(\ref{thetahalf}).  With $\theta$ taken positive, it is also the 
rotation angle from the Breit frame helicity state to the light-front state,
\begin{equation}
	| p,\mu \rangle_L = R_y(\theta) | p,\mu \rangle_B  \ ,
\end{equation}

\noindent which leads to
\begin{equation}
	_L\langle p,\lambda | p,\mu \rangle_B = d^j_{\lambda\mu}(-\theta)
	= d^j_{\mu\lambda}(\theta)\ .
\end{equation}

For the outgoing hadron, the helicity state is (see Fig.~\ref{one} to 
get the angle), 
\begin{equation}
	| p',\mu' \rangle_B = R_y(\pi-\theta) e^{-i K_3 \xi} 
			| rest,\mu' \rangle_z   \ ,
	\end{equation}

\noindent where $K_3$ is the boost operator for the $z$-direction and 
$\xi$ is a rapidity given in terms of the energy $E'$ of the outgoing hadron,
\begin{equation}
	\xi = {\rm arccosh} {E' \over M} = 
		{\rm arccosh} {Q^2 + M^2 + m^2 \over 2mM}  \ .
	\end{equation}

\noindent The light-front state is given by first boosting to the 
correct $(p')^+ = m$ for outgoing hadron (a boost in the negative 
$z$-direction, if $Q \ne 0$), followed by a boost to get the correct 
transverse momentum, which is the same as for the photon.  One has 
\begin{equation}
	| p',\lambda' \rangle_L = e^{ -i Q E_1 / m } e^{ -iK_3 \xi'}
			| rest,\lambda' \rangle_z   \ ,
	\end{equation}

\noindent where 
\begin{equation}
	\xi' = - {\rm arccosh} { M^2 + m^2 \over 2mM}
	\end{equation}

\noindent and $E_1$ is the light-front transverse boost
\begin{equation}
	E_1 = K_1 + J_2  \ .
	\end{equation}


\noindent Thus the overlap is
\begin{eqnarray}
	_B\langle p', \mu' | p',\lambda' \rangle_L &=&
		 _z\langle rest,\mu' | e^{iK_3 \xi} e^{iJ_2 (\pi-\theta)}
		e^{ -i Q E_1 / m } e^{ -iK_3 \xi'}
			| rest,\lambda' \rangle_z  
		\nonumber \\
		&\equiv&  _z\langle rest,\mu' | R_y(-\theta')
			| rest,\lambda' \rangle_z   
		= d^{j'}_{\mu'\lambda'}(-\theta')\ ,
	\end{eqnarray}

\noindent where we know the product of the four operators can only be a 
rotation because the rest four momentum is undisturbed.  Consistent with 
our previous choice, we define $\theta'$ as the angle rotating from the 
rest state connected to the Breit frame helicity state to the 
corresonding state connected to the light-front 
state.  Our method for finding $\theta'$ is to choose a representation 
for the operators, namely 
\begin{equation}        \label{representation}
	J_2 = {1\over 2} \sigma_2 \ , \quad 
	K_3 = {i\over 2} \sigma_3 \ , \quad {\rm and} \quad 
	E_1 = {1\over 2} (i\sigma_1 + \sigma_2) \ ,
	\end{equation}

\noindent where the $\sigma_i$ are the usual $2 \times 2$ Pauli 
matrices, and then to multiply the operators out explicitly.  The result is
\begin{equation}
	\tan \theta' = - { 2 M Q \over Q^2 - M^2 + m^2 } \ ,
	\end{equation}
	
\noindent equivalent to the useful half-angle version given earlier, 
Eq.~(\ref{thetaprime}).

Putting the pieces together gives the light-front to Breit frame 
helicity amplitude conversion formula, quoted in Eq.~(\ref{main}).  The 
inverse of this relation follows using 
\begin{equation}
	d^j_{\mu\lambda_1}(\theta) d^j_{\mu\lambda}(\theta) 
	= \delta_{\lambda_1\lambda}  \ ,
	\end{equation}

\noindent and is
\begin{equation}
	G^\nu_{L\lambda'\lambda} =  d^{j'}_{\lambda'\mu'}(-\theta') \ 
		G^\nu_{B\mu'\mu} \ d^j_{\mu\lambda}(\theta)  \ .
	\end{equation}
	

\section{Light-Front Discrete Symmetry}   \label{discrete}


The discrete symmetries of parity inversion and time reversal are not 
compatible with the light-front requirement that $q^+ = 0$.  However, 
putting all momenta in the $x$-$z$ plane, we can compound the usual 
parity and time reversal operators with 180$^\circ$ rotations about the  
$y$-axis to produce useful and applicable light-front parity and time 
reversal operators~\cite{soper}.


\subsection{Light-front parity}


Let $\mathbb P$ 
be the ordinary unitary 
parity operator that takes 
$\vec x \rightarrow -\vec x$ and $t \rightarrow t$.  Define the 
light-front parity operator by~\cite{soper,jacobwick} 
\begin{equation}
	Y_P = R_y(\pi) \mathbb P.
	\end{equation}

Since $Y_P$ commutes with operators $E_1$ and $K_3$, one has that $Y_P$ 
acting on a light-front state gives 
\begin{eqnarray}
	Y_P |p,\lambda \rangle_L 
	= Y_P e^{-iE_1 p_\perp / p^+} e^{-iK_3 \xi} |rest,\lambda\rangle_z
	= e^{-iE_1 p_\perp / p^+} e^{-iK_3 \xi} Y_P |rest,\lambda\rangle_z
	\ .
	\end{eqnarray}

\noindent Further,
\begin{equation}
	Y_P |rest,\lambda\rangle_z = \eta_P R_y(\pi) |rest,\lambda\rangle_z
	= \eta_P |rest,\lambda'\rangle_z d^j_{\lambda'\lambda}(\pi) \ ,
	\end{equation}

\noindent where $\eta_P$ is the intrinsic parity of the state. Then 
using $d^j_{\lambda'\lambda}(\pi) = (-1)^{j+\lambda} 
\delta_{\lambda',-\lambda}$, one gets for the states 
\begin{equation}
	Y_P |p,\lambda \rangle_L 
	= \eta_P (-1)^{j+\lambda} | p,-\lambda \rangle_L \ .
	\end{equation} 

For current component $J^+$, since $Y_P$ is unitary,
\begin{eqnarray}
	_L\langle p',\lambda' | J^+ | p,\lambda \rangle_L
	&=& \, _L\langle Y_P(p',\lambda') | \ Y_P J^+ Y_P^\dagger \ Y_P
						| p,\lambda \rangle_L
\nonumber \\
	&=& \eta_P^\prime \eta_P (-1)^{j'-j+\lambda'-\lambda} \,
		_L\langle p',-\lambda' | J^+ | p,-\lambda \rangle_L \ .
	\end{eqnarray}

\noindent Hence, the parity relation for light-front helicity amplitudes is
\begin{equation}                      \label{parity}
	G^+_{L,-\lambda',-\lambda} = 
	\eta_P^\prime \eta_P (-1)^{j'-j+\lambda'-\lambda}
	G^+_{L\lambda'\lambda} \ .
	\end{equation}

The parity relation for the usual (Breit frame) helicity ampitudes is 
known~\cite{jacobwick}, and is usually given in terms of amplitudes with 
definite photon helicity, which we define in section~\ref{equivalence}.  
We shall only note that we can derive the relation from the light-front 
result just above, and quote for completeness, 
\begin{equation}
	G^{-\lambda_\gamma}_{B,-\mu',-\mu} = 
	\eta_P^\prime \eta_P (-1)^{j'+j}
	\, G^{\lambda_\gamma}_{B\mu'\mu} \ .
\end{equation}
	

\subsection{Light-front time reversal}


Let $\mathbb T$ 
be the ordinary time 
reversal operator which takes  
$t \rightarrow -t$ and $\vec x \rightarrow \vec x$ and which is 
antiunitary.  By known arguments, time reversal acting on a state at 
rest reverses the spin projection, and one has 
\begin{equation}
	{\mathbb T} | rest, \lambda \rangle_z 
	= (-1)^{j-\lambda} | rest, -\lambda \rangle_z \ .
	\end{equation}

\noindent (By way of review, one starts with 
${\mathbb T} | rest, \lambda \rangle_z 
 = \eta_T(\lambda) | rest, -\lambda \rangle_z$,
and recalls that the states with different $\lambda$ are related by the 
angular momentum raising and lowering operators $J_\pm$.  One shows 
that  $\eta_T(\lambda)$ changes sign as the spin projection changes by 
one unit by considering how $\mathbb T$ commutes with the raising and 
lowering operators.  That only leaves 
$\eta_T(j)$ to be fixed.  Since $\mathbb T$ is antiunitary, one can 
change $\eta_T(j)$ by changing the phase of the state, and one chooses 
the phase of the state so that $\eta_T(j)$ is one.)

Define a light-front time reversal operator by
\begin{equation}
	Y_T = R_y(\pi) \mathbb T \ ,
	\end{equation}

\noindent giving
\begin{equation}
	Y_T | rest, \lambda \rangle_z = | rest, \lambda \rangle_z \ .
	\end{equation}

\noindent This also works for moving light-front states.  Since $Y_T$ 
is  antiunitary, 
\begin{equation}
	Y_T iK_3 Y_T^{-1} = iK_3 \quad {\rm and} \quad 
	Y_T iE_1 Y_T^{-1} = iE_1  \ ,
	\end{equation}

\noindent from which we see,
\begin{equation}
	Y_T | p,\lambda \rangle_L 
	= Y_T \, e^{-iE_1 p_\perp/p^+} 
			e^{-iK_3 \xi} | rest,\lambda \rangle_z
	= | p,\lambda \rangle_L  \ .
	\end{equation}

We use time reversal first to show that the light-front amplitudes are 
real, for current component $J^+$, still remembering that $Y_T$ is antiunitary,
\begin{equation}
	_L\langle p',\lambda' | J^+ | p,\lambda \rangle_L
	= \, _L\langle   Y_T (p',\lambda') | 
		Y_T J^+ Y_T^{-1} Y_T | p,\lambda \rangle_L^*
	= \, _L\langle p',\lambda' | J^+ | p,\lambda \rangle_L^* \ ,
	\end{equation}

\noindent or,
\begin{equation}
	G^+_{L\lambda'\lambda} = ( G^+_{L\lambda'\lambda} )^*
	\end{equation}

\noindent (for momenta in the $x$-$z$ plane).

In general, it is not useful to reverse the initial and final states 
because the particles are different.  But for the elastic case we can 
use further time reversal to relate amplitudes with interchanged 
helicity. First note that for the light-front frame
with the target at rest, the initial and final particles have the same 
$p^+$, so to get a state with the final momentum requires just the 
transverse boost, 
\begin{equation}
	| p',\lambda \rangle_L = e^{-iQE_1/p^+} | p,\lambda \rangle_L \ .
	\end{equation}

\noindent Beginning by applying the previous time reversal result to the 
elastic case, and recalling that $E_1$ commutes  with ``$+$'' components 
of four-vectors, leads to 
\begin{eqnarray}
	G^+_{L\lambda'\lambda}
	&=& \, _L\langle p',\lambda' | J^+ | p,\lambda \rangle_L
	 =  \, _L\langle p,\lambda | J^+ | p',\lambda' \rangle_L
					\nonumber \\
	&=& \, _L\langle p,\lambda | J^+ e^{-iQE_1/p^+}
		| p,\lambda' \rangle_L
	 =  \, _L\langle p,\lambda | e^{-iQE_1/p^+} J^+
		| p,\lambda' \rangle_L
					\nonumber \\
	&=& \, _L\langle p,\lambda | e^{-iJ_3 \pi} e^{+iQE_1/p^+} 
		e^{iJ_3 \pi} 
J^+ | p,\lambda' \rangle_L
					\nonumber \\
	&=& (-1)^{\lambda'-\lambda} \, _L\langle p',\lambda |
				J^+ | p,\lambda' \rangle_L  \ .
	\end{eqnarray}

Thus when the incoming and outgoing particles have the same identity, 
time reversal gives 
\begin{equation}
	G^+_{L\lambda'\lambda} = (-1)^{\lambda'-\lambda}
			G^+_{L\lambda\lambda'}   \ .
	\end{equation}
	
Similarly to the close of the last subsection, we record the time 
reversal result for the helicity amplitudes in the Breit frame, for 
identical incoming and outgoing particles, 
\begin{equation}
	G^{\lambda_\gamma}_{B\mu'\mu} = (-1)^{\mu'-\mu}
			G^{\lambda_\gamma}_{B\mu\mu'}   \ .
	\end{equation}


\subsection{The $x$-Breit frame}


Note that $\theta = -\theta'$ for the equal mass case.  While some of 
the transformations are easy in the target rest frame, where we 
calculated, visualizing this result is not.  For this purpose, the 
$x$-Breit frame, where the incoming and outgoing particles are both 
along the $x$-direction, works well.

The momenta are, in $(p^0,p^1,p^2,p^3)$ notation,
\begin{eqnarray}
p  &=& (\sqrt{m^2 + Q^2/4} ,-Q/2, 0 , 0 ) \ , \nonumber \\
p' &=& (\sqrt{m^2 + Q^2/4} , Q/2, 0 , 0 ) \ , \nonumber \\
q  &=& (     0             , Q  , 0 , 0 ) \ ,
\label{x-Breit-kinematics}
\end{eqnarray}

\noindent and the incoming states are defined by,
\begin{eqnarray}
	|p,\lambda\rangle_L &=& e^{iE_1 Q /2p^+} e^{-iK_3 \xi_1}
				|rest, \lambda \rangle_z \ , \nonumber \\
	|p,\lambda\rangle_B &=& R_y(-\pi/2) e^{-iK_3 \xi}
				|rest, \lambda \rangle_z \ .
\end{eqnarray}

The outgoing states have the same longitudinal boosts, but have opposite 
transformations for getting the transverse momentum.  (The boost 
parameters are not the same.  The transformation with $\xi$ gives a 
momentum along the $z$-direction with the final energy; the 
transformation with $\xi_1$ gives a momentum along the $z$-direction 
with the final $p^+$, but with energy and $p^z$ 
different from the final ones. From the kinematics given by 
Eq.~(\ref{x-Breit-kinematics}), 
we find
$\xi_1 = {\rm arccosh}\frac{m^2 + Q^2/8}{m\sqrt{m^2 + Q^2/4}}$  
and 
$\xi = {\rm arccosh}\frac{\sqrt{m^2 + Q^2/4}}{m}$.
)

Formally, one defines angle $\theta$ from
\begin{equation}
_L\langle p,\lambda | p,\mu \rangle_B = 
	\, _z\langle rest,\lambda | e^{-iJ_2 \theta} | rest, \mu \rangle_z
	\ ,
	\end{equation}

\noindent with a corresponding equation involving the final states and 
angle $\theta'$.  Using the representation given earlier in 
Eq.(~\ref{representation}), 
\begin{equation}
e^{-i\sigma_2 \theta/2} = 
		e^{-\sigma_3 \xi_1/2} e^{(\sigma_1 - i \sigma_2)Q/4p^+}
		e^{i\sigma_2 \pi/4} e^{\sigma_3 \xi/2}  \ .
\end{equation}

\noindent Conjugating the above equation with $\sigma_3$ 
(i.e., taking $\sigma_3 \ldots \sigma_3$) gives
\begin{equation}
e^{+i\sigma_2 \theta/2} = 
		e^{-\sigma_3 \xi_1/2} e^{-(\sigma_1 - i \sigma_2)Q/4p^+}
		e^{-i\sigma_2 \pi/4} e^{\sigma_3 \xi/2}
		= e^{-i\sigma_2 \theta'/2}  \ ,
\end{equation}

\noindent and $\theta = -\theta'$.

\begin{figure}[h]

\epsfig{figure=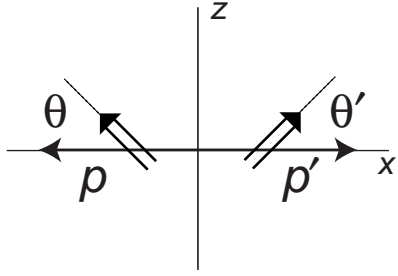}

\caption{Momenta and spin directions for light-front helicity states in 
the $x$-Breit frame.  The momenta are in the $\pm x$-direction and the 
spin directions for the light-front states are indicated by the doubled 
lines.}    \label{xBreit}

\end{figure}

Pictorially, we draw the momenta in Fig.~\ref{xBreit}, and for the 
helicity states the particle spins point along the direction of the 
momenta.  The incoming and outgoing light front states both start with a 
boost in the $z$-direction, and then receive symmetrically opposite 
transverse boosts which rotate the spin vectors in 
opposite directions by the same amount.  The angles $\theta$ and 
$\theta'$ are indicated in the figure.  One can see both that the size 
of the angles should be the same and that the senses should be opposite.


\section{Consequences}      \label{consequences}



\subsection{Light-front parity and the angular condition}


The general angular condition for current component $J^+$ reads
\begin{equation}
	d^{j\prime}_{\lambda'\mu'}(-\theta') \ 
		G^+_{L\lambda'\lambda} \ d^j_{\mu\lambda}(\theta) = 0
		\quad {\rm for} \quad |\mu' + \mu| \ge 2  \ .
	\end{equation}
	
\noindent Say that $\mu+\mu' \ge 2$.  By changing the sign of both $\mu$ 
and $\mu'$ it looks like we could get another angular condition, 
\begin{equation}
	d^{j\prime}_{\lambda',-\mu'}(-\theta') \ 
		G^+_{L\lambda'\lambda} \ d^j_{-\mu,\lambda}(\theta) = 0 \ .
	\end{equation}

\noindent However, using the first of the identities
\begin{equation}                    \label{d_function}
	d^j_{m'm} (\theta) = (-1)^{m-m'} d^j_{-m',-m}(\theta)
	= d^j_{-m,-m'}(\theta)  = (-1)^{m-m'} d^j_{m m'}(\theta)
\end{equation}

\noindent and the light-front parity relation, Eq.(~\ref{parity}), one can 
show by a series of reversible steps that each angular condition with 
$\mu + \mu' \le -2$ is equivalent to one with $\mu + \mu' \ge 2$.  
Hence, we only need to consider cases where $ \mu + \mu' \ge 2$.


\subsection{The angular condition for deuterons}


We shall implement the general angular condition in a couple of special 
cases, rewriting the angular dependence in terms of $Q^2$ and masses.  
For the deuteron, the angular condition comes only from $\mu=\mu'=1$ 
and  we have 
\begin{equation}
d^1_{\lambda' 1}(-\theta') G^+_{L\lambda'\lambda} d^1_{1\lambda}(\theta)
	= 0  \ .
\end{equation}

\noindent For the equal mass case, the arguments of the $d$-functions 
are the same, 
\begin{equation}
	\tan \theta = - \tan  \theta' = {2m_d\over Q}
\end{equation}

\noindent ($m_d$ is the deuteron mass). Using light-front parity, 
Eq.~(\ref{parity}), and the $d$-function identities, 
Eq.~(\ref{d_function}), 
one gets 
\begin{eqnarray}
	G^+_{L++} \left( 
		\left(d^1_{11}\right)^2 + \left(d^1_{1,-1}\right)^2 \right)
	&-& \left( G^+_{L0+}-G^+_{L+0} \right)  d^1_{10}
		\left(d^1_{11} - d^1_{1,-1}\right)
							\nonumber \\
	&+& G^+_{L+-} 2 d^1_{11} d^1_{1,-1}
	- G^+_{L00} \left( d^1_{10} \right)^2  = 0 \ .
\end{eqnarray}

\noindent Substituting for the $d^1$'s and $\tan\theta$, and using the 
light-front time reversal result $G^+_{L+0} = -G^+_{L0+}$, leads to the 
angular condition in its known form~\cite{earlyangular,chh97}, 
\begin{equation}
	(2\eta + 1 ) G^+_{L++} 
	+ \sqrt{8\eta} G^+_{L0+}
	+ G^+_{L+-} - G^+_{L00}  = 0  \ ,
\end{equation}

\noindent where $\eta = Q^2/4m_d^2$.
For the record, we have removed an overall factor, $1/2(1+\eta)$.

Recently, Bakker and Ji~\cite{bakkerji} obtained two constraints on the 
deuteron helicity amplitudes by noting that there were five amplitudes,  
and that all five could be derived from three independent form factors. 
Both constraints they called angular conditions.  They appeared  
differently in different frames; their Drell-Yan-West frame results can 
be most directly compared to our present results.  The constraint they 
call ``AC1'' is, for  momenta in the $x$-$z$ plane, just 
$G^+_{L0+}+G^+_{L+0}=0$. In the present paper this follows from 
light-front time reversal invariance.  
Their constraint ``AC2'' is then precisely the same as the angular 
condition here.


\subsection{A consequence of the angular condition for deuterons}


Perturbative QCD predicts, as we shall review below, that the hadron 
helicity conserving amplitude $G^+_{00}$ is the leading amplitude at
high $Q$ and that
\begin{eqnarray}
G^+_{+0} &=& {a \Lambda_{\rm QCD} \over Q} G^+_{00} \nonumber \\
G^+_{+-} &=& \left( b \Lambda_{\rm QCD} \over Q\right)^2 G^+_{00}
\end{eqnarray}

\noindent to leading order in $1/Q$.  No statement is initially
made about the size of $a$ and $b$.

One may go further, following Chung {\it et al.}~\cite{chung88} or 
Brodsky and Hiller~\cite{bh92} (who interestingly mention 
the work of Carlson and Gross~\cite{cg84} 
in this regard),
to argue that the scale of QCD is  given by $\Lambda_{\rm QCD}$ and that 
we can implement this in the light-front frame by saying that 
\begin{equation}
a,b = O(1) \ .
\end{equation}

\noindent A consequence of this, written in terms of the deuteron 
charge, magnetic and quadrupole form factors~\cite{acg80}, is that to 
good approximation one gets the ``universal ratios''~\cite{bh92}, 
\begin{equation}                    \label{universal}
G_C:G_Q:G_M = \left( {2\over 3}\eta - 1 \right) : 1 : -2  \ .
\end{equation}

\noindent This agrees with the leading power of $Q^2$  
result~\cite{cg84} that $G_C = (2/3)\eta G_Q$, but goes beyond it and 
also gives a prediction for $G_M$.  

We have so far in this subsection used only three light-front helicity 
amplitudes.  There are more that are not zero, and we find a difficulty 
when we  discuss a fourth.  Amplitude $G^+_{++}$ is related to the 
others by the angular condition quoted above.  Also, the perturbative 
QCD arguments that give the scaling behavior of the other helicity 
amplitudes give for $G^+_{++}$ at very high $Q^2$, 
\begin{equation}
G^+_{++} = \left( c \Lambda_{\rm QCD} \over Q\right)^2 G^+_{00} \ .
\end{equation}

\noindent (Helicity is conserved, but other spin dependent 
rules~\cite{cg84} dictate a two power asymptotic suppression of $G^+_{++}$.
This is also consistent with a naturalness condition discussed in 
Ref.~\cite{bakkerji}.
)

The angular condition to leading order now reads,
\begin{equation}
1 + \sqrt{2}\, {a \Lambda_{\rm QCD} \over m_d}  -
 {1\over 2} \left( c \Lambda_{\rm QCD} \over  m_d \right)^2  =  0  \ .
\end{equation}

\noindent The hypothesis that $\Lambda_{\rm QCD}$ sets the scale of the 
subleading amplitudes would suggest that $c$ as well as $a$ is of ${\cal 
O}(1)$. Given the angular condition result just above, this cannot be 
right; at least one of  $a$ and $c$ must
 be $O(m_d/\Lambda_{\rm QCD}) \approx 20$.  Hence the hypothesis is not 
generally workable, and one needs to consider thinking the same about 
the next-to-leading corrections in the ``universal ratios'' expression, 
Eq.~(\ref{universal}).


\subsection{The angular condition for $N$-$\Delta$ transitions}


The $\gamma^* N \rightarrow \Delta$(1232) transition is an important 
reaction that involves final and initial states  with different spins 
and masses.  This makes working out the angular condition more involved 
technically, but not unduly so, as we shall demonstrate. 

There is one angular condition,
\begin{eqnarray}
0 &=& d^{3/2}_{\lambda',3/2}(-\theta') \ G^+_{L\lambda'\lambda} \ 
	  d^{1/2}_{1/2,\lambda }( \theta )
	  						\nonumber \\
  &=& G^+_{L,-3/2,1/2} \left(-d^{3/2}_{3/2,3/2} d^{1/2}_{1/2,-1/2}
  			 + d^{3/2}_{-3/2,3/2} d^{1/2}_{1/2,1/2} \right)
  							\nonumber \\
  &+& G^+_{L,-1/2,1/2} \left( \ d^{3/2}_{1/2,3/2} d^{1/2}_{1/2,-1/2}
  			+ d^{3/2}_{-1/2,3/2} d^{1/2}_{1/2,1/2} \right)
  							\nonumber \\
  &+& G^+_{L,1/2,1/2} \left(-d^{3/2}_{-1/2,3/2} d^{1/2}_{1/2,-1/2}
 			+ d^{3/2}_{ 1/2,3/2} d^{1/2}_{1/2,1/2} \right)
  							\nonumber \\
  &+& G^+_{L,3/2,1/2} \left( \ d^{3/2}_{-3/2,3/2} d^{1/2}_{1/2,-1/2}
  			+ d^{3/2}_{ 3/2,3/2} d^{1/2}_{1/2,1/2} \right) \ ,
\end{eqnarray}

\noindent using the light-front parity.  Explicit substitution for the 
$d$-functions yields 
\begin{eqnarray}
0 = - \cos^2{\theta'\over 2} \sin{\theta'\over 2} \cos{\theta\over 2} 
	&\Bigg\{& G^+_{L,-3/2,1/2}   \left( -\tan{\theta\over 2}  
		\cot{\theta'\over 2} + \tan^2{\theta'\over 2} \right)
  							\nonumber \\
  &-& \sqrt{3} \ G^+_{L,-1/2,1/2}  
  	\left( \tan{\theta\over 2}  +  \tan{\theta'\over 2} \right)
  							\nonumber \\
  &+& \sqrt{3} \ G^+_{L,1/2,1/2} 
  	\left( 1 -  \tan{\theta\over 2} \tan{\theta'\over 2} \right)
  							\nonumber \\
  &-& G^+_{L,3/2,1/2} \left( \cot{\theta'\over 2} + 
 		 \tan{\theta\over 2} \tan^2{\theta'\over 2} \right)
  			\Bigg\}\ .
\end{eqnarray}

\noindent  Finally, removing the overall factors and substituting for 
the trigonometric functions gives the angular condition for the 
$N$-$\Delta$ transition, 
\begin{eqnarray}
0  &=& \left[ (M-m)(M^2-m^2) + mQ^2 \right] G^+_{L,-3/2,1/2}
	+ \sqrt{3} M Q (M-m) G^+_{L,-1/2,1/2}
				\nonumber \\
   &+& \sqrt{3} M Q^2     G^+_{L,1/2,1/2}
	+ Q \left[ Q^2 -m(M-m) \right]  G^+_{L,3/2,1/2},
\end{eqnarray}

\noindent where $m$ is the nucleon mass and $M$ is the $\Delta$ mass.  
For the record, we have removed another overall factor,
$Q_+ (Q_+ - Q_-) / (2m M^2 Q^2)$.

The asymptotic scaling rules, cited in the next subsection, say that 
$G^+_{L,1/2,1/2}$ goes like $1/Q^4$ at high $Q$, that $G^+_{L,3/2,1/2}$ 
and $G^+_{L,-1/2,1/2}$ go like $1/Q^5$, and that $G^+_{L,-3/2,1/2}$ goes 
like $1/Q^6$.  If we write 
\begin{equation}
G^+_{L,3/2,1/2} =  {b \Lambda_{\rm QCD} \over Q} G^+_{L,1/2,1/2}
\end{equation}

\noindent modulo logarithms at high $Q$, then the leading $Q$ part of 
the angular condition says 
\begin{equation}
\sqrt{3} +  {b \Lambda_{\rm QCD} \over M} = 0 \ .
\end{equation}


\subsection{Equivalence of leading powers in Breit and Light-front 
frames}                                     \label{equivalence}


The idea of ``good currents'' and ``bad currents'' is native to the 
light-front frame.  In analyzing the power law scaling behavior at high 
$Q^2$, for a given helicity amplitude, it is often thought to be safest 
to stay in the light-front frame and use only ``good currents.''  We 
shall here derive the Breit frame helicity amplitude scaling behaviors 
>from their light-front counterparts.  
Note here $q^+ = 0$ both in the light-front frame and the Breit frame 
that we discuss in this subsection.
All the $q^+ = 0$ frames are related to each other only by the 
kinematical operators that make the light-front time $\tau$ intact.
We will find, nicely enough, that 
the scaling behaviors are the same as one would have found using the 
Breit frame only.
That is, one can get the correct leading power scaling behavior from a 
Breit frame analysis alone.

For a light-front helicity amplitude, the scaling behavior at high $Q$ is 
\begin{equation}
G^+_{L\lambda'\lambda} \propto \left( m\over Q \right)
^{2(n-1) + |\lambda'-\lambda_{min}| + |\lambda-\lambda_{min}|}
		\ ,
\end{equation}

\noindent where $n$ is the number of quarks in the state, $m$ is a mass 
scale, and $\lambda_{min}$ is the minimum helicity of the incoming or 
outgoing state (i.e., 0 or 1/2 for bosons or fermions, respectively).

Regarding the Breit frame, we have thus far given its helicity  
amplitudes in terms of $G^\nu_B$ where $\nu$ is a Lorentz index.  It is 
usual to substitute a photon helicity index for the Lorentz index, 
using (for incoming photons)
\begin{equation}
G^{\lambda_\gamma}_{B\mu'\mu} = \epsilon_\nu(q,\lambda_\gamma) \,
		G^\nu_{B\mu'\mu} \ ,
\end{equation}

\noindent with polarizations (in $(t,x,y,z)$-type notation)
\begin{eqnarray}
\epsilon_\pm &=& \epsilon(q,\lambda_\gamma=\pm) = 
		(0,\pm\cos\theta,-i,\pm\sin\theta)/\sqrt{2}
							\nonumber \\
\epsilon_0   &=& \epsilon(q,\lambda_\gamma=0  ) = 
		(\csc\theta,\cos\theta,0,-\cos^2 \theta \csc\theta) \ ,
\end{eqnarray}

\noindent where $\theta$ is the angle between $\vec q$ and the negative 
$z$-direction, as shown in Fig.~\ref{one}.  One can work out that 
\begin{equation}
\eta \equiv (1,0,0,-1) = \sin\theta \left[\epsilon_0 -
 {1\over\sqrt{2}} \left( \epsilon_+ - \epsilon_- \right)
			\right] \ ,
\end{equation}

\noindent so that
\begin{equation}
G^{\nu=+}_{B..} = \sin\theta \left[ G^0_{B..} -{1\over\sqrt{2}} \left( 
G^+_{B..} - G^-_{B..} \right)
	\right] \ ,
\end{equation}

\noindent where the superscripts on $G_B$ will in the rest of this 
subsection refer to photon helicity unless explicitly stated otherwise.  
With the general relation, Eq.~(\ref{main}), this gives directly the 
expression we use to obtain the scaling behavior of the Breit amplitudes,
\begin{equation}						
\label{expression}
G^0_{B\mu'\mu} - {1\over\sqrt{2}} \left( G^+_{B\mu'\mu} 
	- G^-_{B\mu'\mu} \right)
	= \csc\theta \  d^{j'}_{\lambda'\mu'}(-\theta') \ 
		G^+_{L\lambda'\lambda} \ d^j_{\mu\lambda}(\theta)  \ .
\end{equation}

\noindent We can select terms on the left-hand-side by choice of $\mu'$ 
and $\mu$, since Breit amplitudes are non-zero only for 
$\lambda_\gamma = \mu' + \mu$.

The $d$-functions can be written in terms of sines and cosines of half 
angles, so we record that at high $Q$, 
\begin{eqnarray}
\sin{\theta\over 2} &=& {m\over Q} + {\cal O}\left( m\over Q \right)^3
	\quad {\rm and} \quad
	\cos{\theta\over 2} = 1 + {\cal O}\left( m\over Q \right)^2
\nonumber  \\
\sin{\theta'\over 2} &=& -{M\over Q} 
+ {\cal O}\left( m\over Q \right)^3
	\quad {\rm and} \quad
 	\cos{\theta'\over 2} = 1 + {\cal O}\left( m\over Q \right)^2 \ ,
\end{eqnarray}

\noindent where $m$ inside the $\cal O$ symbol is a generic mass scale.  
The $d$-functions can be expanded as~\cite{edmonds} 
\begin{eqnarray}
d^j_{\mu\lambda}(\theta) &=& a_1 
		\left( \cos{\theta\over 2} \right)^{2j - |\mu-\lambda|}
		\left( \sin{\theta\over 2} \right)^{|\mu-\lambda|}
						\nonumber \\
&+& a_2 \left( \cos{\theta\over 2} \right)^{2j - |\mu-\lambda|-2}
		\left( \sin{\theta\over 2} \right)^{|\mu-\lambda|+2} + \ldots,
\end{eqnarray}

\noindent where $a_1,a_2,\ldots$ are numerical coefficients with 
$a_1 \ne 0$.  Thus for large $Q$,
\begin{equation}
d^j_{\mu\lambda}(\theta) \propto
		\left( m \over Q \right)^{|\mu-\lambda|} 
		+ {\cal O}\left( m\over Q \right)^{|\mu-\lambda| + 2} \ .
\end{equation}

On the right-hand-side of Eq.~(\ref{expression}), there is no term that 
falls slower than the term that has $\lambda'=\lambda=\lambda_{min}$, 
and $G^+_{L,\lambda_{min},\lambda_{min}} \propto (m/Q)^{2(n-1)}$.  Thus 
the Breit amplitude leading falloff at high $Q$ is
\begin{equation}
G^{\lambda_\gamma}_{B\mu'\mu} \propto \left( m\over Q \right)^
		{-1 + 2(n-1) + |\mu'-\lambda_{min}| + |\mu-\lambda_{min}|} \ .
\end{equation}

\noindent These are the same results one can get by directly analyzing 
amplitudes in the Breit frame for various photon helicities~\cite{breitresult}.

By way of example, we will give the Breit and light-front frame helicity 
amplitudes for elastic electron-nucleon scattering.  In terms of the 
standard Dirac, Pauli, and Sachs form factors one may work out 
\begin{eqnarray}
G^+_{L++} &\equiv& \, _L\langle p',{1\over 2} \Big| J^+ \Big| p,{1\over 
2} \rangle_L = 2 p^+ F_1(q^2) \ , \nonumber \\
G^+_{L-+} &\equiv& \, _L\langle p',-{1\over 2} \Big| J^+ \Big| p,{1\over 
2}\rangle_L = 2 p^+ {Q\over 2m} F_2(q^2) \ ,
\end{eqnarray}

\noindent for the light-front states and
\begin{eqnarray}
G^+_{B++} &\equiv& \, _B\langle p',{1\over 2} \Big| \epsilon_+ \cdot J 
\Big| p,{1\over 2} \rangle_B = \sqrt{2}  Q G_M(q^2) \ ,\nonumber \\
G^0_{B-+} &\equiv& \, _B\langle p',-{1\over 2} \Big| \epsilon_0 \cdot J 
\Big| p,{1\over 2} \rangle_B =  2m G_E(q^2) \ ,
\end{eqnarray}

\noindent for the Breit frame states.  The scaling rules predict that 
$G_M$, $G_E$, and $F_1$ scale as $1/Q^4$ and that $F_2$ scales as 
$1/Q^6$, consistent with the relations 
\begin{eqnarray}
G_M &=& F_1 + F_2   \ ,   \nonumber \\
G_E &=& F_1 + {q^2 \over 4m^2} F_2  \ .
\end{eqnarray}

\noindent (There is recent data~\cite{gayou} and 
commentary~\cite{ralstonetal} on the helicity flip scaling results.)


\section{Summary and Conclusion}


The purpose of the present paper has been largely kinematical.  We have 
examined the relationship between the helicity amplitudes in the Breit 
and light-front frames.  One particular result has been a clear view of 
where the angular condition comes from. 
The angular condition is a constraint on light-front helicity 
amplitudes.  It follows from applying angular momentum conservation in 
the Breit frame, where the application of angular momentum conservation 
to the helicity amplitudes is elementary.  One consequence is that an 
amplitude must be zero if it requires the photon to have more than one 
unit magnitude of helicity, and this statement cast in terms of 
light-front amplitudes is the angular condition~\cite{capstick95,earlyangular}.

Another set of constraints follows from parity and time-reversal 
invariance.  Neither of  these symmetries can be used directly on the 
light-front because the light-front has a preferred spatial direction.  
However, each of them can be modified to give a valid symmetry operation 
(at least for strong and electromagnetic interactions) for the 
light-front~\cite{soper}.  To define a light-front parity, choose the 
$x$-$z$ plane to contain all the momenta and then consider mirror 
reflection in the $x$-$z$ plane.  This reflection leaves the momenta 
unchanged but reverses the helicities~\cite{jacobwick}.  Technically, it 
is the same as ordinary parity followed by a 180$^\circ$ rotation about 
the $y$-axis, and helicity amplitude relations that follow from it were 
given in section~\ref{discrete}.

Similarly one defines light-front time reversal as ordinary time reversal 
followed by the 180$^\circ$ rotaton about the $y$-axis.  Time reversal 
invariance implies that the amplitudes are always real for momenta in 
the $x$-$z$ plane, with additional relations possible for elastic 
scattering, as detailed also in section~\ref{discrete}.

The general angular condition appears compactly in terms of 
$d$-functions, the representations of the rotation operators.  It can be 
rewritten in terms of masses and momentum transfer.  We gave the 
translations for two cases.  For electron-deuteron elastic scattering, 
the result is well known~\cite{earlyangular}.  Nonetheless, it does have 
an unchronicled (we believe) consequence.  That is that the mass scale 
associated with the asymptotic power-law falloff of non-leading 
amplitudes must generally be of order of the nucleon or deuteron mass.  
It had been hoped that the light-front was a favored frame where the 
non-leading amplitudes would have small numerical coefficients:  
asymtotically of order $\Lambda_{\rm QCD}/Q$, to an appropriate power, 
times the leading amplitude.  As the angular condition contradicts this, 
it also takes away the motivation that underlay the suggestion of the 
universal behavior for spin-1 form factors.

Additionally, we gave the angular condition explicitly for the 
$N$-$\Delta$ electromagnetic transition.  We believe this result is also 
new.  The angular conditon here fixes precisely the leading power of one 
subleading amplitude. 

The power law scaling of the helicity amplitudes can be analyzed, with a 
definite power of $1/Q$ given in terms of the number of constituents in 
the wave function and in terms of the helicities of the incoming and 
outgoing states, and this can be done in 
either the light-front frame or in the Breit frame.  A preference for 
one or the other is sometimes  given.  Our final ``application'' was to 
obtain the Breit frame scaling from the light-front scaling using the 
transformation between them given earlier in section~\ref{relatons}, and 
to find that the result is the same as one obtains doing the analysis 
directly in the Breit frame.

\begin{acknowledgments}
CEC thanks the DOE for support under contract DE-AC05-84ER40150, under 
which the Southeastern Universities Research Association (SURA) operates 
the Thomas Jefferson National Accelerator Facility, and also thanks the 
NSF for support under Grant PHY-9900657.  
CRJ thanks the DOE for support under contract DE-FG02-96ER40947 and the 
NSF for support under Grant INT-9906384, and also thanks the North 
Carolina Supercomputing Center and the National Energy Research Scientific
Computer Center for the grant of supercomputing time.
\end{acknowledgments}


\noindent


\begin{thebibliography}{99}

\bibitem{Dirac} P.A.M. Dirac, Rev. Mod. Phys. {\bf 21}, 392 (1949).

\bibitem{Steinhardt} P.J. Steinhardt, Ann. Phys. {\bf 128}, 425 (1980).

\bibitem{lcca} S.Fubini, G. Furlan, Physics {\bf 1}, 229 (1965);
S. Weinberg, Phys. Rev. {\bf 150}, 1313 (1966);
J. Jersak and J. Stern, Nucl. Phys. {\bf B7}, 413 (1968);
H. Leutwyler, in Springer Tracts in Modern Physics Vol. 50,
ed. G. H\"ohler, (Berlin, 1969).

\bibitem{lcparton} J.D. Bjorken, Phys. Rev. {\bf 179}, 1547 (1969);
S.D. Drell, D. Levy, T.M. Yan, Phys. Rev. {\bf 187}, 2159 (1969), 
{\it ibid.} \bf D1\rm, 1035 (1970).

\bibitem{BPP} S.J. Brodsky, H.C. Pauli, and S.S. Pinsky, 
Phys. Rept. {\bf 301}, 299 (1998).

\bibitem{Ja2} W. Jaus, Phys. Rev. {\bf D44}, 2851 (1991).

\bibitem{CJ1} H.-M. Choi and C.-R. Ji, Phys. Rev. {\bf D59}, 074015 (1999);
Phys. Rev. {\bf D56}, 6010 (1997).

\bibitem{Kaon} H.-M. Choi and C.-R. Ji, Phys. Lett. {\bf B460}, 461 (1999);
Phys. Rev. {\bf D59}, 034001 (1999).

\bibitem{DYW} S.D. Drell and T.M. Yan, Phys. Rev. Lett. {\bf 24} (1970) 181;
\\ G. West, Phys. Rev. Lett. {\bf 24} (1970)

\bibitem{Ji1} H.-M. Choi and C.-R. Ji, Phys. Rev. {\bf D58}, 071901 (1998).

\bibitem{chad} C.-R.Ji and C. Mitchell, Phys. Rev. {\bf D62}, 085020 (2000).

\bibitem{jacobwick}
M.~Jacob and G.~C.~Wick,
Annals Phys.\  {\bf 7}, 404 (1959)
[Annals Phys.\  {\bf 281}, 774 (2000)].

\bibitem{wick62}
G.~C.~Wick,
Annals Phys.\  {\bf 18}, 65 (1962).

\bibitem{capstick95}
S.~Capstick and B.~D.~Keister,
Phys.\ Rev.\ D {\bf 51}, 3598 (1995)
[arXiv:nucl-th/9411016].


\bibitem{soper}
D.~E.~Soper,
Phys.\ Rev.\ D {\bf 5}, 1956 (1972).

\bibitem{earlyangular} 
H.~Osborn,
Nucl.\ Phys.\ B {\bf 38}, 429 (1972);
H.~Leutwyler and J.~Stern,
Annals Phys.\  {\bf 112}, 94 (1978);
I.~L.~Grach and L.~A.~Kondratyuk,
Sov.\ J.\ Nucl.\ Phys.\  {\bf 39}, 198 (1984)
[Yad.\ Fiz.\  {\bf 39}, 316 (1984)];
B.~D.~Keister,
Phys.\ Rev.\ D {\bf 49}, 1500 (1994)
[arXiv:hep-ph/9303264].

\bibitem{chh97}
C.~E.~Carlson, J.~R.~Hiller and R.~J.~Holt,
Ann.\ Rev.\ Nucl.\ Part.\ Sci.\  {\bf 47}, 395 (1997).

\bibitem{bakkerji}
B.~L.~Bakker and C.~R.~Ji,
Phys.\ Rev.\ D {\bf 65}, 073002 (2002)
[arXiv:hep-ph/0109005].


\bibitem{chung88}
P.~L.~Chung, W.~N.~Polyzou, F.~Coester and B.~D.~Keister,
Phys.\ Rev.\ C {\bf 37}, 2000 (1988).

\bibitem{bh92}
S.~J.~Brodsky and J.~R.~Hiller,
Phys.\ Rev.\ D {\bf 46}, 2141 (1992).

\bibitem{cg84}
C.~E.~Carlson and F.~Gross,
Phys.\ Rev.\ Lett.\  {\bf 53}, 127 (1984).

\bibitem{acg80} See for example
R.~G.~Arnold, C.~E.~Carlson and F.~Gross,
Phys.\ Rev.\ C {\bf 21}, 1426 (1980).

\bibitem{edmonds} 
A.R. Edmonds, {\it Angular Momentum in Quantum Mechanics},
Princeton University Press, 1957.

\bibitem{breitresult} The Breit frame result can be gotten using 
techniques shown in~\cite{cg84}.


\bibitem{gayou}
O.~Gayou {\it et al.}  [Jefferson Lab Hall A Collaboration],
=  5.6-GeV**2,'' Phys.\ Rev.\ Lett.\  {\bf 88}, 092301 (2002)
[arXiv:nucl-ex/0111010];
M.~K.~Jones {\it et al.}  [Jefferson Lab Hall A Collaboration],
p(pol.),'' Phys.\ Rev.\ Lett.\  {\bf 84}, 1398 (2000)
[arXiv:nucl-ex/9910005].


\bibitem{ralstonetal}
J.~P.~Ralston and P.~Jain,
arXiv:hep-ph/0207129;
J.~P.~Ralston, P.~Jain and R.~V.~Buniy,
AIP Conf.\ Proc.\  {\bf 549}, 302 (2000)
[arXiv:hep-ph/0206074];
G.~A.~Miller and M.~R.~Frank,
Phys.\ Rev.\ C {\bf 65}, 065205 (2002)
[arXiv:nucl-th/0201021];
S.~J.~Brodsky,  Talk at Workshop on Exclusive Processes at High Momentum 
Transfer, Newport News, Virginia, 15-18 May 2002, 
arXiv:hep-ph/0208158;
A.~V.~Belitsky, X.~Ji and F.~Yuan,
arXiv:hep-ph/0212351.




\end{thebibliography}
\end{document}